\documentclass[11pt,twoside]{article}
\usepackage{newpasp}

\usepackage{epsf}

\index{Duc, P.--A.}
\index{Brinks, E.}

\begin{document}

\title{H\,{\footnotesize I} Recycling: Formation of Tidal Dwarf Galaxies}
\author{P.--A. Duc}
\affil{CNRS and CEA--Saclay, Service d'astrophysique, Orme des
Merisiers, 91191 Gif sur Yvette cedex, France} 
\author{E. Brinks}
\affil{Departamento de Astronom\'{\i}a, Apdo. Postal 144, Guanajuato,
 Gto. 36000, Mexico}



\begin{abstract}

Galactic collisions trigger a number of phenomena, such as
transportation inward of gas from distances of up to kiloparsecs from
the center of a galaxy to the nuclear region, fuelling a central
starburst or nuclear activity. The inverse process, the ejection of
material into the intergalactic medium by tidal forces, is another
important aspect and can be studied especially well through detailed
H\,{\sc i} observations of interacting systems which have shown that a
large fraction of the gaseous component of colliding galaxies can be
expelled. Part of this tidal debris might fall back, be dispersed
throughout the intergalactic medium or recondense to form a new
generation of galaxies: the so--called tidal dwarf galaxies. The
latter are nearby examples of galaxies in formation.  The properties
of these recycled objects are reviewed here and different ways to
identify them are reviewed.

\end{abstract}




\section{Recycling the H\,{\footnotesize I} gas}

The VLA and other synthesis arrays have, for twenty years and more,
revealed the spectacular distribution of the atomic hydrogen (H\,{\sc i})
in interacting systems.  Long tidal tails that are even more prominent
than their optical counterparts, bridges, and ring--like structures
are among the many weird and wonderful features which show up in high
resolution maps (see Schiminovich et al., this volume).  Because in
general the H\,{\sc i} in disk galaxies extends well beyond the optical
$R_{25}$ radius, this atomic hydrogen reacts very efficiently to any
external perturbation as evidenced by the fact that most of the neutral
gas is found outside the colliding disks. In systems like Arp~105 and
NGC~7252 (see Fig.~1), up to 90\% of the gas visible at 21~cm is
situated in the intergalactic medium (IGM) and little, if any gas remains
within the parent spirals. In addition to gas being actively removed, a large 
part of the original H\,{\sc i} has likely been funnelled inward where it
was transformed into another
phase (see review by Struck, 1999).

The fate of the tidally expelled gas will largely depend on its
density and location with respect to the interacting galaxies. Whereas
the clouds closest to the interacting (or merging) pair will fall back
at timescales of a few Myr (Hibbard \& Mihos, 1995), the most distant
ones will become gravitationally unbound.  They might slowly diffuse
and enrich the IGM with heavy elements or the stars might even form
the basis of the faint background light in the intracluster medium
(ICM) if the interacting system belongs to a cluster. Under certain
conditions, however, tidal debris may be recycled. If self--gravity is
sufficiently large, expelled clouds will condense and collapse again
to build new star--forming objects. Offspring as massive as magellanic
dwarf irregular galaxies has been observed around several interacting
systems (see for examples Fig.~1). Generally situated at the tip of
50--100 kpc long tidal tails, they are referred to as Tidal Dwarf
Galaxies (TDGs).  The total gas fraction that could end up in a TDG is
yet largely unknown and it is hoped that numerical simulations might
one day get a handle on this. Observationally, H\,{\sc i} clouds towards
TDGs as massive as $5~10^{9}\,{\rm M_\odot}$ have been measured.

\begin{figure}
\plotone{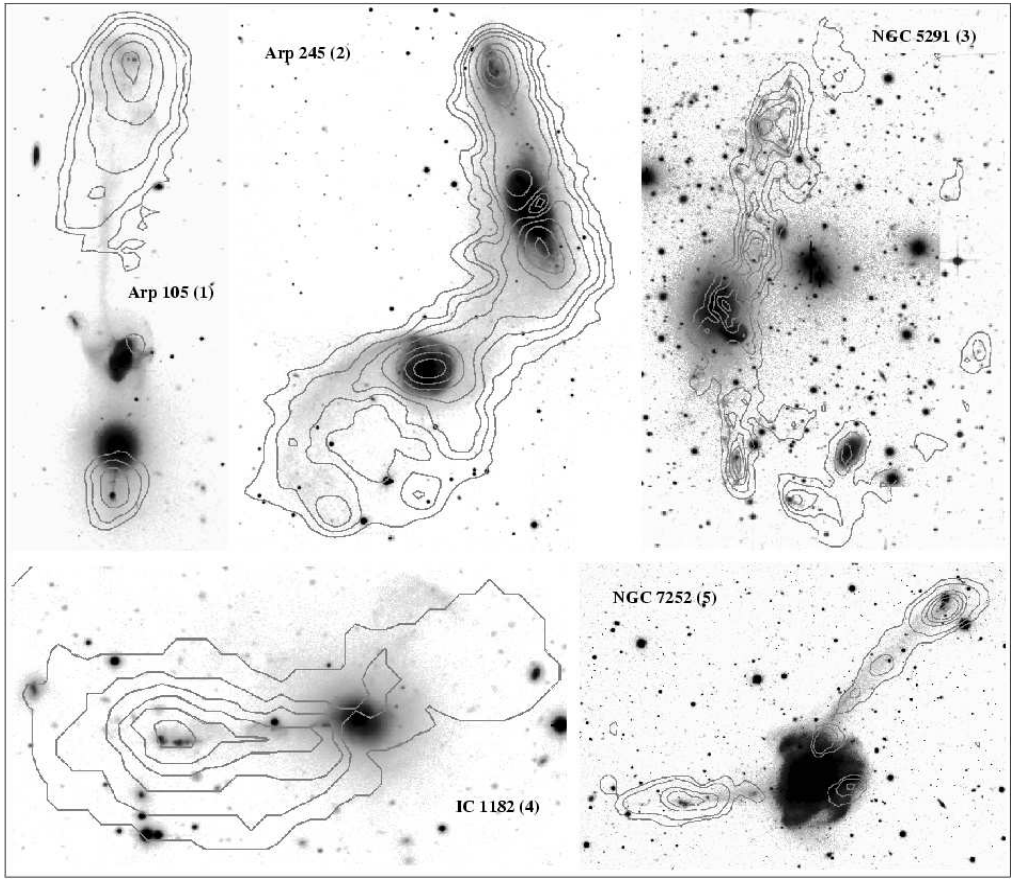}
\caption{\small H\,{\sc i} distributions for a sample of interacting
systems which have formed some tidal dwarf galaxies. The H\,{\sc i}
contours are superimposed on optical $V$--band images;
references for the H\,{\sc i} VLA data: 1) Duc et al.\ (1997); 2) Duc et
al.\ (2000); 3) Malphrus et al.\ (1997), and Duc \& Mirabel (1998); 4)
Dickey (1997); 5) Hibbard et al.\ (1994)}
\label{fig1}
\end{figure}

\section{Forming tidal dwarf galaxies}

Not only H\,{\sc i} clouds are expelled during tidal interactions, but
any material that was originally (rotating) in a disk, in particular
the stars. Therefore TDGs are mixed bags, composed of young stars
formed {\em in situ} from collapsing gas clouds and an older
population pulled out from their parent galaxies. The latter component
may be unimportant in systems involving gas--rich early type galaxies
(i.e. NGC~5291, Duc \& Mirabel 1998; see also Fig.~1). 
Instabilities in the gaseous --- H\,{\sc i} --- component seem to be the
driving factor in the formation of TDGs.
  Star formation occurs at rates which might
reach 0.1\,M$_\odot\,{\rm yr^{-1}}$.  Given that TDGs contain huge
H\,{\sc i} reservoirs, typically $10^9\,{\rm M_\odot}$, it is to be
expected that the relative importance of an older stellar population
will decrease with time as SF proceeds.
Appreciable quantities of molecular gas have also recently been
detected in TDGs by Braine et al.\ (2000). They suggest that this gas
has been formed {\em in situ} out of the collapsed H\,{\sc i}.

\section{Identifying tidal dwarf galaxies}

Tidal objects will survive provided that they have a potential well
that is deep enough to sustain themselves against internal or external
disruption. When identifying TDGs, it is therefore important to
check that they are not simply the agglomerated debris of a collision
but are self--gravitating entities (Duc et al.\ 2000). One should
hence try to distinguish those tidal features that are kinematically
decoupled from their host tails, the kinematics of which is governed
by streaming motions. Because of the difficulty of obtaining high
sensitivity, high resolution H\,{\sc i} data, and problems related to
projection effects along the line of sight, this is a difficult task
that requires a careful examination of the available datacubes, such
as those provided by synthesis radio observations (for the neutral gas
component), and Fabry--Perot or any integral field instrument (for the
ionised component). Addressing the stellar kinematics would be even
more challenging.  So far, evidence for such self--gravitating clouds
have been found in the interacting systems Arp~105 (Duc et al.\ 1997;
see Fig.~2), NGC~5291 (Duc \& Mirabel 1998) and perhaps Arp~245 (Duc
et al.\ 2000).

\begin{figure}
\plotone{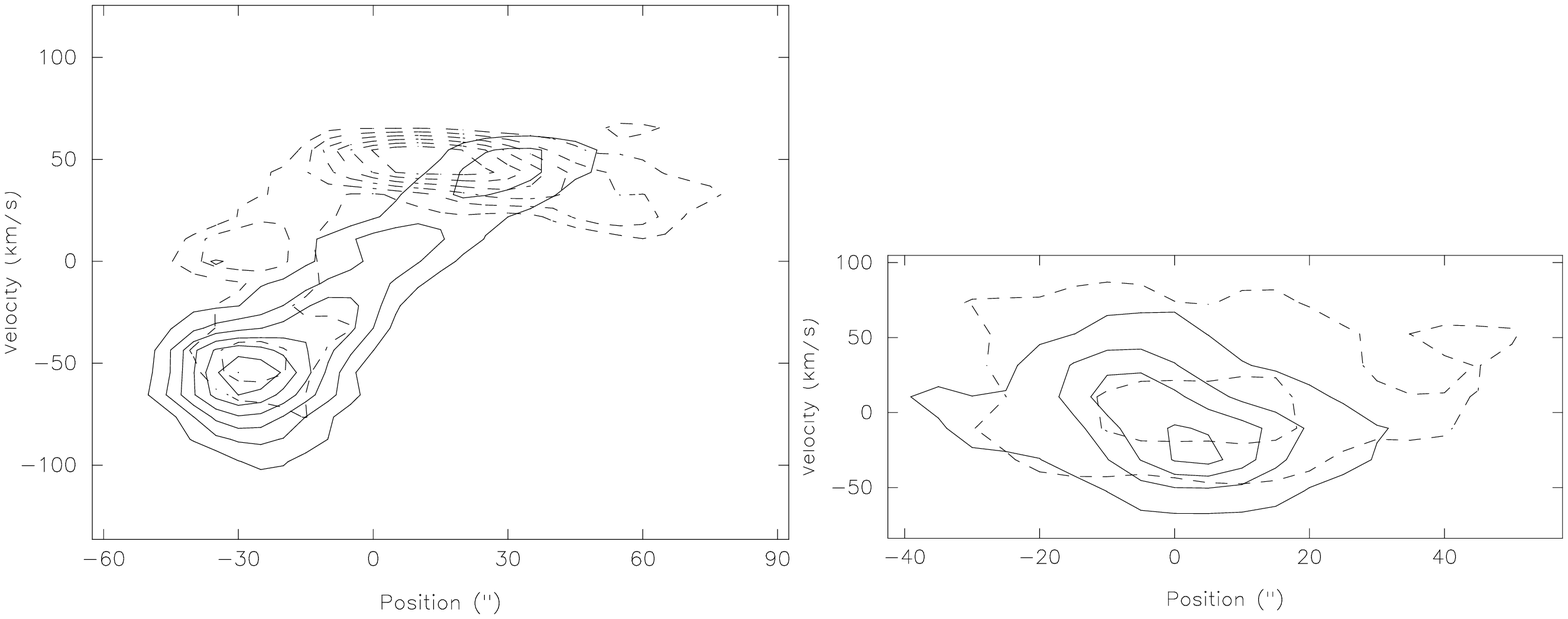}
\caption{\small H\,{\sc i} position--velocity diagrams towards the tidal
tails of Arp 105 (left) and Arp 245 (right). Two components have been
identified and disentangled using cuts made at different positions
along the tails: one component is associated with gas following a
streaming motion along the tail (dashed contours), and a second one
which appears to be kinematically decoupled (full contours) and which
coincides spatially with the optical tidal dwarf galaxy. Adapted from
Duc et al.\ 1997, and Duc et al.\ 2000.}
\label{fig2}
\end{figure} 

It is one thing to pick out TDGs which are still linked to the tidal
tails out of which they formed. But how can one recognise old tidal
dwarf galaxies that would have lost their physical connection with
their parent galaxies as the tidal tail linking it vanishes with time?
One might make use of three special properties of TDGs. First of all,
their metallicity: due to the fact that they are recycled material, it is 
much higher than that of classical dwarf galaxies of the same
luminosity (Duc et al.\ 2000). The oxygen abundance of recently formed
TDGs averages out at one third of solar, i.e., the abundance of
spirals at or slightly beyond the optical $R_{25}$ radius.  Secondly,
it is expected that TDGs contain little if any dark matter {\em if} DM
in their progenitors was distributed in a large halo, as is
traditionally assumed (Barnes \& Hernquist, 1992).  If however, DM is
present in disks, for instance in the form of cold molecular gas
(Pfenniger et al., this volume) DM dominated tidal tails should
form and TDGs would have a substantial DM content. Finally, the
stellar population of TDGs should at least be bimodal: a fraction of
their stars has originally come from the parent disks, whereas another
part has been formed {\em in situ}. Reconstructing the star formation
history of a galaxy using for instance its color--magnitude diagram
might reveal a tidal origin.
These methods have already been used to identify some older TDG
candidates, both in the field as well as in clusters (Hunter et al.\
2000; Duc et al., in prep). But so far, the overall fraction of
dwarf galaxies of tidal origin remains unknown.



\acknowledgements 
We are grateful to our many collaborators involved
in this multiwavelength study, and in particular to Felix Mirabel,
Volker Springel, Barbara Pichardo, Peter Weilbacher and Jonathan Braine.  We thank the organisers
 for a very successful and pleasant conference.

\end{document}